\documentclass[sn-mathphys]{sn-jnl}
\usepackage{natbib}
\usepackage{graphicx}%
\usepackage{multirow}%
\usepackage{amsmath,amssymb,amsfonts}%
\usepackage{amsthm}%
\usepackage{mathrsfs}%
\usepackage[title]{appendix}%
\usepackage{xcolor}%
\usepackage{textcomp}%
\usepackage{manyfoot}%
\usepackage{booktabs}%
\usepackage{algorithm}%
\usepackage{algorithmicx}%
\usepackage{algpseudocode}%
\usepackage{listings}%
\usepackage{caption}
\usepackage{subcaption}
\usepackage[utf8]{inputenc}
\begin{document}

\title[Article Title]{Secure Information Embedding in Images with Hybrid Firefly Algorithm}


% \author*[1,2]{\fnm{First} \sur{Author}}\email{iauthor@gmail.com}



% \affil*[1]{\orgdiv{Department}, \orgname{Organization}, \orgaddress{\street{Street}, \city{City}, \postcode{100190}, \state{State}, \country{Country}}}

% \affil[2]{\orgdiv{Department}, \orgname{Organization}, \orgaddress{\street{Street}, \city{City}, \postcode{10587}, \state{State}, \country{Country}}}

% \affil[3]{\orgdiv{Department}, \orgname{Organization}, \orgaddress{\street{Street}, \city{City}, \postcode{610101}, \state{State}, \country{Country}}}

\author*[1]{\fnm{Sahil} \sur{Nokhwal}}\email{snokhwal@memphis.edu}

\author[2]{\fnm{Manoj} \sur{Chandrasekharan}}\email{manoj.c@memphis.edu}

\author[3]{\fnm{Ankit} \sur{Chaudhary}}
\email{dr.ankit@ieee.org}


\affil*[1]{\orgdiv{Department of Computer Science}, \orgname{University of Memphis}, \orgaddress{\city{Memphis}, \state{TN}, \country{USA}}}

\affil[2]{\orgdiv{Department of Electrical and Computer Engineering}, \orgname{University of Memphis}, \orgaddress{\city{Memphis}, \state{TN}, \country{USA}}}

\affil[3]{\orgdiv{Department of Computer Science}, \orgname{University of Missouri}, \orgaddress{ \city{St Louis}, \state{Missouri}, \country{USA}}}


\abstract{
Various methods have been proposed to secure access to sensitive information over time, such as the many cryptographic methods in use to facilitate secure communications on the internet. But other methods like steganography have been overlooked which may be more suitable in cases where the act of transmission of sensitive information itself should remain a secret. Multiple techniques that are commonly discussed for such scenarios suffer from low capacity and high distortion in the output signal.
This research introduces a novel steganographic approach for concealing a confidential portable document format (PDF) document within a host image by employing the Hybrid Firefly algorithm (HFA) proposed to select the pixel arrangement. This algorithm combines two widely used optimization algorithms to improve their performance. The suggested methodology utilizes the HFA algorithm to conduct a search for optimal pixel placements in the spatial domain. The purpose of this search is to accomplish two main goals: increasing the host image's capacity and reducing distortion. Moreover, the proposed approach intends to reduce the time required for the embedding procedure. The findings indicate a decrease in image distortion and an accelerated rate of convergence in the search process. The resultant embeddings exhibit robustness against steganalytic assaults, hence rendering the identification of the embedded data a formidable undertaking.
}

\keywords{Metaheuristic optimization, Image steganography, Data hiding, Machine Learning, Information security, Steganography}

\footnotetext[1]{ * 
% {\fontspec{Symbola}\symbol{"2709}} 
\text{ Corresponding author: Sahil Nokhwal (snokhwal@memphis.edu)}}
% \footnotetext[3]{* / ✉\text{ Corresponding author: Ankit Chaudhary dr.ankit@ieee.org}}


\maketitle

\section{Introduction}\label{sec:intro}
The increasing adoption of the World Wide Web has led to the emergence of several individuals engaged in cybercriminal operations. These individuals aggressively search for and exploit vulnerabilities in vital communication infrastructure in order to hijack or disrupt them for the goal of sabotage or financial extortion. This has become an even more pressing issue with the rise of hacker groups that are incentivized and backed by nation-states that can execute attacks at a scale that wasn't possible earlier. One of the most common forms of such attacks is gaining access to secure networks like corporate or government networks to steal sensitive information that is routinely shared on such networks. If communications on such systems are not sufficiently secured, it results in successful attacks leading to major disruptions in their day-to-day operations. So use of highly secure encryption in communications is a critical part of any network.

To guarantee the secure transfer of data over a network, it is essential to deploy suitable security measures that effectively protect vital information from illegal interception. Numerous encryption techniques have been suggested over the years to achieve this objective, whereby the source system encrypts the data prior to transmission, and afterward, it is decrypted on the recipient's system. This measure safeguards the data against unauthorized access and guarantees the integrity of the received data. Although such data encryption systems have been a standard in network communication for a long time, there are still some use cases where such systems can be improved. 

Humans have a well-known psychological trait of assigning a greater value to objects they can't own/access and hence perceive a higher cost of access to be justified even if the underlying object doesn't really have such high value. This scarcity mentality is regularly exploited by marketing agencies to sell luxury products at very high prices. But this same principle is also one of the main problems with encryption as a method of security. Any instance of encrypted communication available in public that can be viewed by unauthorized parties and this can act as a temptation for some to try and gain access to such information, even if the risk of legal action or the cost of actually breaking the encryption is too high. The mere presence of a cipher motivates at least some people to try and break the cipher and read its contents. Given enough technical skill and computational power, no encryption is 100\% safe, so the safer method of security is to hide information in plain sight. Steganographic techniques do just that and ensure that sensitive data remains far from prying eyes.

Two important types of information concealment that have been covered in the past are watermarking and steganography. Each has certain benefits and drawbacks. The process of watermarking involves the encoding of transmissions with a discernible symbol, such as embedding a signature within signals. This serves the purpose of authorizing the owner of the signals and confirming their ownership. In general, a diminutive sign consisting of a range from one to several thousand bits is employed. The principal objective of steganography is to achieve clandestine communication, with the intention of concealing the presence of a message from an external observer. In contrast, cryptography does not endeavor to obscure the existence of covert communication; rather, its objective is to make a message indecipherable to an unauthorized entity. In order to achieve such outcomes, the desired information is embedded in a host signal (usually an audio or visual signal). The large size of host data allows reasonable bits of desired information to be hidden in between them without raising any suspicions. This methodology can be utilized across various domains, including the authentication of ownership, ensuring resistance against tampering, and facilitating the secure transport of confidential data.

Most cases where data security is critical involve some kind of text document like intelligence assessments, patent applications or even emails and other private communications between journalists and their sources. PDF documents, being one of the most universally used document formats, it is important for any steganographic method to ensure that even large files like these can be secured with little delay and with minimum distortion.

In this paper, we discuss the use of steganography to hide a PDF document in an image. This document may include any text data and may also include embedded images. In order to protect it against attacks by illicit actors, we wish to safeguard the file from public detection. In order to reduce visual distortion, we employed a meta-heuristic method called the Hybrid Firefly algorithm (HFA) developed by  \cite{zhang2016novel}, to determine the best locations for concealing data. This requires comparing every combination of pixel values, which might take an exponentially long time to compute. This process was expedited by the use of the HFA algorithm.
This work encompasses the following contributions: 
%contributions
\begin{enumerate}
    \item This paper presents a unique steganographic approach that seeks to embed a confidential PDF file inside a specified host image.
    \item Extensive experiments have demonstrated minimal distortion when using different host images.
\end{enumerate}

This study introduces a new approach to improve the output of steganography by using the Hybrid Firefly algorithm. This work is structured in the ensuing fashion. Section \ref{sec:relworks} lays out the previous work done in this field and the current state of steganographic techniques. The existing degree of progress in the Firefly algorithm (FA) in \cite{yang2009firefly} and Differential Evolution (DE) algorithm in \cite{storn1997differential} has been thoroughly investigated in Section \ref{ssec:firefly} and Section \ref{ssec:dealgo}, including its implementation. In Section \ref{ssec:hfirefly}, the proposed methodology and its resulting consequences are outlined. The utilization of the HFA algorithm involves the development of a comprehensive approach that encompasses LSB substitution and the selection of optimal pixels. Section \ref{sec:results} presents the experimental results, accompanied by a visual comparison of alternative methodologies that are available. Section \ref{sec:conclusion} presents a concise summary of the contributions demonstrated in the earlier sections.

\section{Related work in context} \label{sec:relworks}
Over the last twenty years, several steganographic applications have emerged, with many utilizing data concealing techniques based on the least-significant-bit (LSB) paradigm. These approaches entail identifying certain pixels that exhibit the necessary qualities inside a host medium. Subsequently, data is integrated into the least significant bit (LSB) of these discovered pixels \cite{adelson1990digital, yadav2016secure, van1994digital, chaudhary2012hash, melman2023comparative}. The authors of the study \cite{wang2001image} proposed the use of a Genetic Algorithm (GA) to determine the most effective replacement matrix for hiding secret messages inside the important section of the cover image. Furthermore, the method of local pixel adjustment (LPAP) was proposed as a means to enhance the quality of the steganographic image. An effective strategy for k-LSB substitution was developed to address the problem presented by a high value of k.

The steganographic technique proposed by \cite{li2007steganographic} employs the JPEG format and Particle Swarm Optimization (PSO). This methodology exhibits potential applications within the spatial domain for transform domain applications. \cite{bedi2011using} suggested the application of PSO for selecting optimal pixels for replacement in a grayscale cover image, suitable for embedding concealed grayscale image pixel data. While the efficacy of this strategy is recognized, it is noteworthy that search algorithms like Cuckoo Search (CS) have demonstrated superior results due to their improved solution space, as evidenced by \cite{nokhwal2023embau}.

Various spatial-domain embedding approaches leverage metaheuristic algorithms for optimization, enhancing the efficiency of the embedding process. Recent steganography works have been proposed by \cite{alkhliwi2023huffman, rathika2023ensemble, kiran2023novel, melman2023efficient, fofanah2023watermarking, yang2023novel, salim2023image, sargunam2023empirical, hameed2023secure, mahalakshmi2023improving, bahaddad2023image, sharma2023optimized, apau2023multilayered, yang2023novel}. In the study by \cite{wazirali2019optimized}, a spatial-domain image steganography system utilizing the LSB procedure is presented. Genetic algorithms (GAs) are utilized to optimize the sequencing of phases in the embedding process, such as pixel analysis, pixel modification, toggling of hidden bits, and other related tasks. The Peak signal-to-noise ratio (PSNR) is used as the aim function to ensure the effective arrangement of message bits within the cover picture. Other neural network-based works are presented in \cite{nokhwal2023pbes, nokhwal2023dss, nokhwal2023rtra, tanwer2020system}.

\cite{chen2021multiple} propose a bidirectional embedding strategy that combines histogram flipping and GA approaches. The researchers analyze the use of histogram flipping for different embedding, focusing on enhancing embedding speed and reducing distortion. The goal function is determined by calculating the discrepancy between maximum and minimum distortion values.

\cite{dougan2016new, tong2023chaotic} leverage chaos maps to enhance the data concealment scheme based on GA. Chaotic maps, including logistic and Gaussian maps, introduce unpredictability to genetic variation. PSNR is used as a fitness measure.

Particle Swarm Optimization (PSO) is a commonly used optimization technique in the field of computer intelligence. The study conducted by \cite{li2018steganography} employs Particle Swarm Optimization (PSO) to improve the effectiveness of pixel-value differencing. The optimization approach chooses the most favorable gray values for each pixel from a set of options provided by the modulus function. The primary goal function selects suitable pixels, whereas the secondary objective function calculates the best solution based on the results of the primary target function.

Utilizing PSO, as demonstrated by \cite{rustad2022optimization}, allows for a significant increase in the embedding capacity within the spatial region of pictures. The Particle Swarm Optimization (PSO) algorithm is utilized to determine the optimal locations for embedding, while optimization techniques are performed to determine the optimum initial implantation point and the most efficient path for scanning pixels.

The work undertaken by \cite{mohsin2019new} utilizes the Artificial Bee Colony (ABC) approach to enhance the allocation of blocks for covert image embedding. This technology exhibits superior resistance to particular types of noise assaults in comparison to similar steganographic techniques.

The solution presented by \cite{bala2022approach} seeks to attain content anonymity for healthcare photographs through the application of medical information masking techniques. This technique utilizes Sudoku-based encryption to guarantee the confidentiality of healthcare photos and exploits the Queen Traversal pattern to identify specific pixels inside the image.

Metaheuristic methods provide substantial advantages in the domains of finance, operational research, and manufacturing. The study undertaken by \cite{nipanikar2018sparse} shows that these domains often include complex challenges that traditional methods struggle to address, particularly in situations with limited access to significant computational resources.

The encryption technique presented by \cite{snasel2020jpeg} uses the Cipher Block Chaining method and is specially designed to process several images simultaneously. Furthermore, it is capable of being used in conjunction with concurrent computing approaches.

The work conducted by \cite{soto2018adaptive} employs the Adaptive Black Hole Algorithm to tackle the set-covering issue, resulting in many globally optimum solutions for different set-covering scenarios. \cite{cheung2016nonhomogeneous} proposes a modification to the CS algorithm to enhance its search capabilities. To enhance visual fidelity, \cite{gerami2012least} uses PSO in combination with Optical Pixel Adjustment. \cite{mohamed2011data} introduces the concept of permutable keys, utilizing it to propose the k-LSB replacement technique. Gene Expressing Programming is introduced by \cite{mohamed2018hiding} for determining the optimal key permutation for LSB replacement. \cite{douiri2017steganographic} proposes a novel technique for steganography in grayscale images by employing graph coloring. An alternative approach is suggested by \cite{habibi2011survey}, introducing a technique for concealing text messages within 24-bit RGB color graphics, employing the Shuffled-Frog-Leaping algorithm. However, none of these works efficiently disguise a PDF file within an image.

\section{Proposed architecture} \label{sec:proparch}

The suggested model's overall structure is founded upon the integration of three distinct algorithms, which together contribute to the attainment of the final model. These algorithms are:
\begin{enumerate}
    \item Firefly algorithm
    \item Differential Evolution algorithm
    \item Hybrid Firefly Algorithm
\end{enumerate}
The following subsections provide a comprehensive analysis of the algorithms' intricacies.
\subsection{Firefly Algorithm} \label{ssec:firefly}
The FA, or Firefly Algorithm, is a meta-heuristic optimization technique developed by Xin-She Yang in 2008. It is inspired by the bioluminescent display demonstrated by tropical fireflies.
The standard Firefly algorithm utilizes the observable communication patterns seen in tropical fireflies and integrates the idealized behavior linked with their flashing patterns. The construction of the mathematical model in FA is based on three idealized principles.
\begin{enumerate}
    \item Fireflies exhibit a unisex characteristic, whereby their attraction towards other fireflies is not contingent upon their respective sexes.
    \item The level of attractiveness exhibited by an individual is directly correlated with their level of brightness. Additionally, it can be inferred that attractiveness diminishes as the distance between individuals rises. Hence, in the case of two fireflies exhibiting intermittent luminescence, it can be observed that the firefly with lower luminosity will exhibit a tendency to travel towards the firefly with higher luminosity.
    \item The luminosity of a firefly is contingent upon the topography underlying the cost function. Hence, within the framework of a maximizing problem, there exists a direct correlation between the amount of light and the value of the fitness function.
\end{enumerate}

The conventional firefly algorithm encompasses two crucial aspects that necessitate consideration: the formulation of the brightness and the alteration in appeal. Initially, it is reasonable to argue that the luminosity of the firefly is contingent upon the cost function landscape. Next, we establish the measurement of light intensity variation and develop a model to quantify the corresponding alteration in attractiveness. It is well-established that the intensity of light exhibits a negative correlation with the distance separating the light source and the medium that absorbs the light. In our computer model, we use the assumption that the brightness $I$ changes exponentially and monotonically with both the distance $r$ and the absorption of light parameter $\gamma$. The aforementioned relationship is expressed as:
\begin{equation}
    I = I_0 \cdot e^{-\gamma r^2}   \label{eq1}
\end{equation}

The symbol $I_0$ represents the initial light intensity emitted from the source, specifically at the point where $r$ = 0. Meanwhile, the symbol $\gamma$ denotes the coefficient that characterizes the absorption of light. Based on the aforementioned idealized principles, it can be inferred that the perceived appealing factor of a firefly is directly correlated with the intensity of its emitted light, denoted as I. The attractive coefficient $\beta$ of the firefly's light can be defined in a manner analogous to the illumination coefficient $I$.

\begin{equation}
    \beta = \beta_0 \cdot e^{-\gamma r^2}   \label{eq2}
\end{equation}

where $\beta_0$ represents the initial level of light attraction when the distance $r$ is equal to zero.
The Euclidean distance is employed to compute the separation between two fireflies, denoted as $i$ and $j$, located at positions $x_i$ and $x_j$ respectively.

\begin{equation}
    c_{ij} = ||x_i-x_j||_2 = \sqrt{\sum_{m=1}^{d}(x_{i, m} - x_{j, m})^{2}}   \label{eq3}
\end{equation}

where $d$ represents the total number of dimensions. The displacement exhibited by firefly $i$ towards firefly $j$, which is more appealing due to its increased luminosity, is ascertained by:

\begin{equation}
    x = x_{i}+ \beta_{0} e^{-\gamma c^2} (x_{j} - x_{i}) + \eta \epsilon_{i}  \label{eq4}
\end{equation}

The initial term in the equation represents the present position of firefly $i$. The subsequent term accounts for the lure between firefly $i$ and firefly $j$. Finally, in the last term, randomization is included by using a vector of independent variables denoted as $\epsilon_i$. These variables are sampled from many distributions, including the Uniform distribution, and the Gaussian distribution. The magnitude of the step size is determined by the coefficient scaling parameter $\eta$ in the third term.

\begin{algorithm}
\caption{Firefly algorithm pseudo-code}\label{alg:FA_algo}
\begin{algorithmic}[1]
\State Objective function $g(a),a=[a_1, ..., a_D]^T$\;
\State Commence the establishment of the firefly population $a_i(i=1,...,n)$\;
\State Determine the magnitude of light intensity $I_i$ at $a_i$ by $g(a_i)$\;
\State A definition of the extinction coefficient $\gamma$\;
\While{$p < TotalGeneration$}

    \For{$i=1:n$ all $n$ fireflies}

        \For{$j=1:n$ all $n$ fireflies}

            \State Determine the Cartesian distance, denoted as $r$, between the points $a_i$ and $a_j$
            
            \If{$I_j > I_i$}
            
                \State Distance influences attractiveness in accordance with the equation \eqref{eq1}\;
                \State Move firefly $i$ towards $j$ in all d dimensions\;
            \EndIf
            \State Assess novel solutions and revise illumination levels\;
        \EndFor
    \EndFor
    \State Rank fireflies and identify the finest individual at the moment\;
\EndWhile
\State Visualize and post-process outcomes\;
\end{algorithmic}
\end{algorithm}

\subsection{Differential Evolution Algorithm} \label{ssec:dealgo}
This algorithm finds an optimal solution for a problem while evolving from some randomly initiated starting points using a vectorized mutation operator combined with either exponential or binomial crossover. It was proposed by \cite{storn1997differential} and there have been many other variants of this algorithm that have been proposed since. One of the most widely used variants is the DE/rand/1/bin or the classic DE variant.
In the context of a particular minimization problem with $d$ dimensions, population consisting of $n$ independent result vectors is utilized. A change in vector $v_i$ is formally characterized as:

\begin{equation}
    v_{i, g+1} = x_{r_{1}, g} + F(x_{r_{2}, g} - x_{r_{3}, g})
\end{equation}

where $r_{1}, r_{2}, r_{1} \epsilon [1, n]$ and $r_{1} \neq r_{2} \neq r_{3} \neq i$ are three randomly selected solutions from the population and g is the generation index. There is also a perturbation parameter $F(F\epsilon [0,2])$ that adjusts the amplification of the difference vector $(x_{r_{2}, g} - x_{r_{3}, g})$.

Similarly, the binomial crossover operation is also used to generate a new trial vector  from the perturbed or mutated vector $v_{i, g+1}$ and the target vector $x_{i, g}$

\begin{equation}
    u_{i,(g+1)} =   \begin{cases}
                        v_{ij, (g+1)}, \hfill \text{ if } r(j) \leq C_r \text{ or } j = random(i)\\
                        x_{ij, g}, \hfill \text{ if } r(j) > C_r \text{ or } j \neq random(i)\
                    \end{cases}
\end{equation}

where the crossover constant $C_{r} \epsilon [0,1]$ and $random \epsilon [1,2,..., d]$ is a random permutation vector index that ensures at least one mutated vector is included in the trailing vector.

The selection mechanism for this algorithm is as below:

\begin{equation}
    x_{i,(g+1)} =   \begin{cases}
                        u_{ij, (g+1)}, \hfill \text{ if } f(u_{i, (g+1)}) \leq f(x_{i, g})\\
                        x_{i, g},  \hfill otherwise\
                    \end{cases}
\end{equation}

This approach has a resemblance to previous algorithms in which a greedy acceptance strategy is used, wherein an update is deemed acceptable only if it leads to an improvement in the present target being pursued.

\begin{algorithm}
\caption{Differential Evolution algorithm pseudo-code}\label{alg:DE_algo}
\begin{algorithmic}[1]

    \State A randomly generating beginning points, i.e., population $x_i(i=1,...,n)$ is initialized\;
    \State Input values for $F$ and $C_r$, represent the crossover probability and perturbation parameters, respectively\;

    \While{$p < TotalGeneration$}

        \For{$x_i$ where $i=1:n$ all $n$ individuals}
        
            \State Randomly select 3 vectors $x_{r_1}$, $x_{r_2}$ and $x_{r_3}$\;
            \State Create a completely novel vector via mutation $v_i$\;
            \State Randomize the index $i$ using $random(i)$ \;
            \State Produce a value that is distributed at random. $r(j)\epsilon [0,1]$\;
            \For{$j=1:D$ all $D$}
            
                \State Determine the Crossover procedure for every parameter by  $v_{ij}$\;
                \begin{equation*}
                    u_{i,(g+1)} = \begin{cases}
                                    v_{ij, (g+1)}\\
                                    x_{ij, g} \
                                \end{cases}
                \end{equation*}
            \EndFor
        \EndFor
    \EndWhile
   
    \State Post process results and visualize\;

\end{algorithmic}
\end{algorithm}

% \begin{algorithm}
% \caption{Differential Evolution algorithm pseudo-code}\label{alg:DE_algo}

% Initialize the population $x_i(i=1,...,n)$with randomly generated starting points\;
% Set the perturbation parameter $F$ and crossover probability parameter $C_r$\;
% \While{$t < MaxGeneration$}
% {
%     \For{$i=1:n$ all $n$ individuals}
%     {
%         For each $x_i$, randomly select 3 vectors $x_{r_1}$,$x_{r_2}$ and $x_{r_3}$ \;
%         Use mutation to generate a new vector $v_i$\;
%         Generate a random index random($i$)\;
%         Generate a randomly distributed value $r(j) \epsilon [0,1]$\;
%         \For{$j=1:D$ all $D$ }
%         {
%             Calculate the Crossover operation for each parameter $v_{ij}$ update\;
%             \begin{equation*}
%                 u_{i,(g+1)} = \begin{cases}
%                                 v_{ij, (g+1)}\\
%                                 x_{ij, g} \
%                             \end{cases}
%             \end{equation*}
%         } 
%     }
% }
% Post process results and visualize\;
% \end{algorithm}

\subsection{Hybrid Firefly Algorithm} \label{ssec:hfirefly}
Both the firefly method and differential evolution provide distinct benefits and demonstrate efficacy across a diverse array of optimization tasks. The hybrid Firefly method is derived from the FA and the DE algorithm. This hybrid approach aims to use the strengths of both algorithms, resulting in an improved optimization technique. The proposed approach integrates the attraction mechanism used in the FA with the diversification mechanism of DE algorithm. This integration aims to enhance the convergence rate while simultaneously preserving population diversity.

The use of an intensification or exploitation strategy involves directing the search towards a specific local location. This direction is based on either past knowledge or newly acquired information obtained during the search process, which suggests the potential presence of an optimal solution inside that particular region. Differential Evolution (DE) is capable of performing localized searches, particularly when converging towards local optimum solutions. Enhancing the accuracy and convergence rate of an algorithm's solution may be achieved by effectively balancing the conflicting needs associated with these two factors. By integrating these two aspects, we may use their respective strengths to improve the efficacy of the hybrid algorithm in both exploitation and exploration.

It is vital to acknowledge that the consolidation and restructuring of individual location data takes place alone after the first iteration of concurrent FA and DE procedures. This stands in opposition to the generation of novel sites by random walks or alternative operators. The primary advantage of using this technique is to prioritize the exploration of current locations inside advantageous regions obtained during the preceding iteration, as opposed to traversing less promising places within the search space.

The hybrid approach demonstrates linear time complexity while the population size remains relatively small. Consequently, it efficiently computes solutions in terms of computational cost, mostly attributed to the evaluation of objective function values.

% declaration of the new block
\algblock{Parallel}{EndParallel}
% customising the new block
\algnewcommand\algorithmicparallel{\textbf{parallel}}
\algnewcommand\algorithmicpardo{\textbf{}}
\algnewcommand\algorithmicendparallel{\textbf{end\ parallel}}
\algrenewtext{Parallel}[1]{\algorithmicparallel\ #1\ \algorithmicpardo}
\algrenewtext{EndParallel}{\algorithmicendparallel}
\begin{algorithm}
% \SetKwBlock{Parallel}{do in parallel}{end}
\caption{Hybrid Firefly algorithm pseudo-code}\label{alg:HFA_algo}
\begin{algorithmic}
    \State The whole group will be partitioned into two distinct groups, denoted as G1 and G2\;
    \State Populations of G1 and G2 are initialized\;
    \State Assess the cost value of every individual\;
    \While{Termination condition unmet}

        \Parallel
            \State Implement the FA procedure on G1 \;
            \State Execute the DE procedure on G2\;
        \EndParallel
            \State Enhance the global optimum throughout the whole of the population\;
            \State Combine and arbitrarily divide categories into the following: Both G1 and G2\;
            \State Assess the fitness quality of every individual.\;
    \EndWhile
    \State Visualize and post-process outcomes\;
\end{algorithmic}

\end{algorithm}

\begin{figure}[htbp]
    \centering    
    \includegraphics[width=\linewidth]{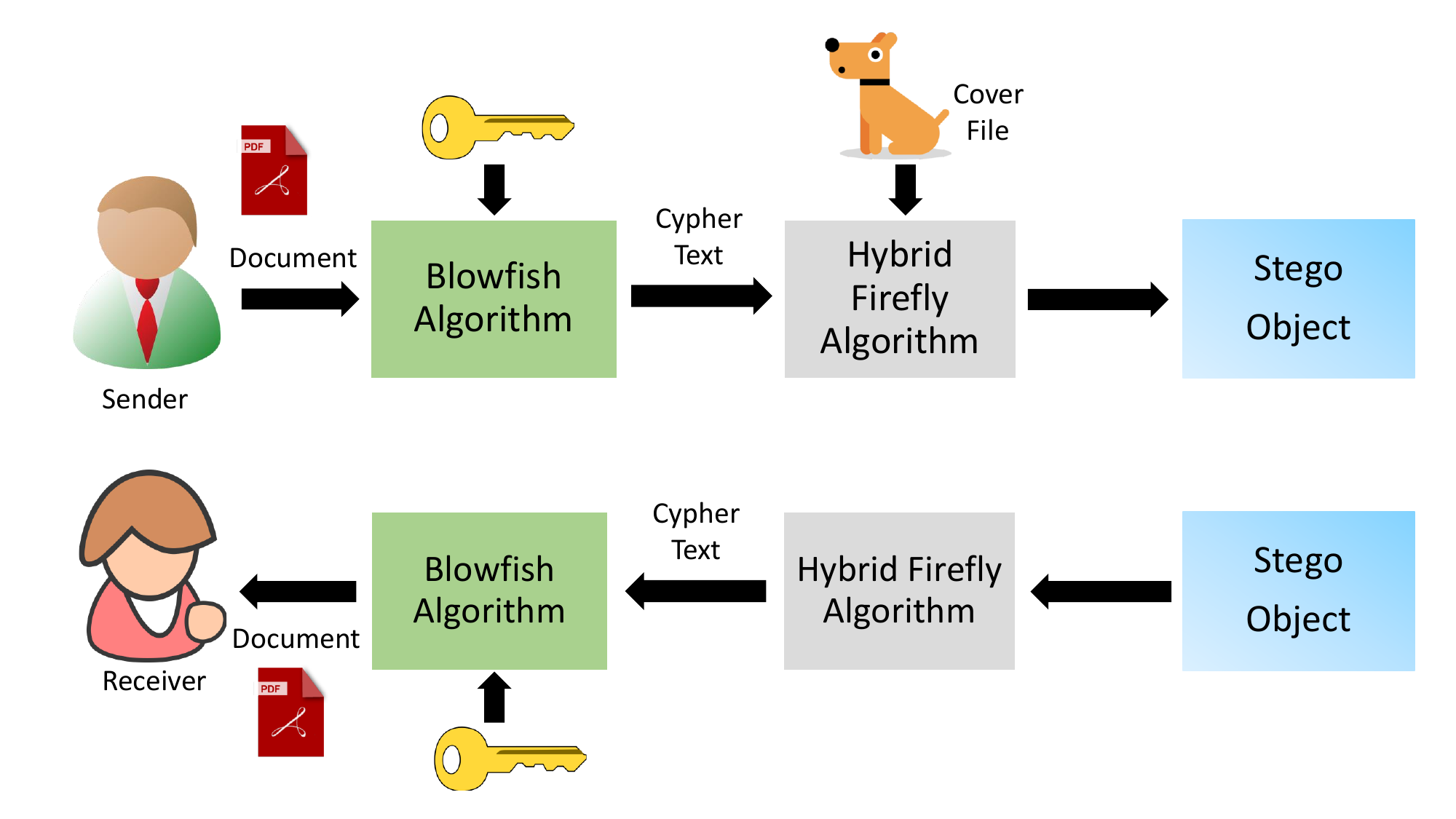}
    \caption{The architectural framework of the suggested steganographic approach}
\end{figure}

\section{Proposed Technique}\label{sec:proposed_tech} 
The proposed approach has two unique elements: the first component includes the embedding of data from PDF documents with the cover picture, while the subsequent component involves the retrieval of concealed information via the steganographic method.

\begin{table}[!b]
 \caption{Algorithm dependent parameters}
    \centering
    \begin{tabular}{|c|c|c|c|c|}
        \hline
         & \multicolumn{4}{|c|}{\textbf{Parameters}}\\
        \hline
        \textbf{FA} & $\beta_{0}=2 \cdot rand$ & $\gamma=1/S^2$ & $\eta=\eta_0 \cdot 0.95^{iter}$ & $\epsilon_{i}=$ Lévy flight\\
        \hline
        \textbf{DE} & \multicolumn{2}{|c|}{$F = 0.5$} & \multicolumn{2}{|c|}{$C_r = 0.9$}\\
        \hline
        % \botrule
    \end{tabular}
    \vspace{0.5mm}   \label{tab:algo_param_table}
\end{table}

\subsection{Embedding document using the HFA algorithm}\label{sec:embedding}
The primary aim of our proposed technique is to determine the optimal arrangement of pixels for the reason of incorporating PDF data into a picture, with the ultimate goal of minimizing any possible alteration in the cover photo. The proposed technique is supported by a structure of procedures that encompasses the subsequent steps:
\begin{enumerate}
    \item The first stage is the perusal of a confidential document intended for incorporation into the cover picture.
    \item The binary numbers derived from the PDF document will undergo encryption using the Blowfish encryption technique, as described by \cite{schneier1993description}.
    \item To ensure the covert embedding of the encrypted information inside the cover picture, the 128-bit encrypted information is converted into a sequence of binary numbers.
    \item The pixels of the cover picture undergo a process of transformation whereby they are converted into their respective 8-bit counterparts.
    \item The objective function is obtained by means of a computational procedure from the multi-objective function. This metric functions as a measure of the degree of accuracy between a steganographic picture and the fundamental data. The calculation of the PSNR entails evaluating the difference between the highest achievable magnitude of the steganographic signal and the strength of the noise that impacts the accuracy of its depiction.
 
    \item \emph{Population initialization:} The firefly population initially forms through a random initialization process in designated groups denoted as $G_1$ and $G_2$. Within the solution space $\Omega \in \mathbb{R}^{d}$, a set of fireflies, denoted as $\mathbb{F}$, is created randomly in each group $\mathbb{F}(1)$, $\mathbb{F}(2)$, $\mathbb{F}(3)$, $\ldots$, $\mathbb{F}(n)$. Here, $\mathcal{D}$ signifies the cumulative determinants. Consequently, we define a vector $\mathbb{F}(i) = \mathbb{F}(2)$, $\mathbb{F}(3)$, $\ldots$, $\mathbb{F}(n)$ to represent the values of the selection variables associated with the $i^{th}$ firefly.

    % \item \emph{Initialize memeplexes:} The number of frogs $\mathcal{F}$, is divided into memeplexes, i.e., $m$ and each memeplex has $\eta$ frogs, where $\eta$ is an integer. Hence, the equation depicting the aggregate number of samples $\mathcal{F}$ in the swamp can be formulated as
    % \begin{equation*}
    %   \mathcal{F} = m \times \eta
    % \end{equation*}
      
    % Here $\mathcal{F}$ depicts the total number of frogs in the initial population, $m$ represents the total number of memeplexes, and $\eta$ corresponds to the number of frogs within every memeplex.
     
    \item \emph{Rank fireflies:} The fitness of every firefly within the set $\mathbb{F}(i)$ is assessed based on the value $fitness\mathbb{(F}(i))$. Subsequently, this procedure is employed to classify the population into distinct clusters. These clusters are then organized in descending order and distributed in a round-robin fashion.
    \item \emph{Execute DE algorithm:} Once the population for group $G_2$ has been initialized and the parameters have been set, the next step is to execute mutation and crossover operations on the potential solutions.
    \item \emph{Rank fitness:} The fitness of each produced vector would be assessed in a manner akin to the firefly method. 
    \item \emph{Perform Search:} An evaluation is carried out inside each group, wherein the proficiency of each participant in each group is assessed. The global results are then updated by considering the fittest individuals from each group. The groups are then merged and subsequently partitioned into the subsequent set of groups denoted as $G_1$ and $G_2$, using a random allocation process.
    Consequently, a new population is formed. The aforementioned stages are iteratively executed until either the specified termination criteria are satisfied or an optimum solution is attained.
    
    \item Utilizing the HFA methodology, ascertain the optimal selection of pixels for the purpose of data hiding inside the cover photo.
    \item The 1-LSB method is used to discreetly include encrypted information bits into a host image, minimizing any noticeable effects on the image's integrity.
\end{enumerate}

\subsection{Extraction of embedded steganographic PDF information} \label{ssec:extration}
\begin{enumerate}
    \item Reduce the dimensionality of a 3-channel steganographic color picture (m x n x 3) to a 2-dimensional image (m' x n').
    \item Equation~\ref{eq:objFun} is used to create a specialized objective function that takes into account SSIM and PSNR values.
    \item \emph{Population initialization:} The initial population is generated in accordance with section~\ref{sec:embedding}
    \item After configuring all parameters, the HFA algorithm is executed in order to minimize the cost function. The result of this procedure is the generation of an ideal arrangement of pixels that could be used for the retrieval of embedded data.
    \item The encrypted data may be accessed by using the order derived from the preceding step.
    \item For decryption, Blowfish algorithm is used to transform the data into its binary representation.
    \item From the binary bits reconstruct a PDF file.
\end{enumerate}

\subsection{Evaluation Metrics} \label{ssec:metrics}
The Peak Signal-to-Noise Ratio (PSNR) is a numerical measure, expressed in decibels (dB), that enables the evaluation of the extent of degradation in the steganographic signal. The correlation between the quality of steganographic photos and the rise in PSNR value is directly proportional.

The SSIM, or Structural Similarity Index, is a quantitative metric utilized to assess the degree of resemblance between two given images. The comparison approach combines the structural characteristics and luminosity measurements of the images under scrutiny. The Structural Similarity Index (SSIM) is calculated by assessing the resemblance between two photographs based on their brightness, contrast, and general arrangement. The system produces a numerical result that is between -1 and 1. A score of 1 indicates a significant degree of resemblance between the two photos, whereas a score of -1 indicates a considerable degree of dissimilarity. 

The Mean Squared Error (MSE) is a widely utilized statistical metric in academics for quantifying the average squared discrepancy between the anticipated and actual values of a designated variable. Currently, the assessment focuses on evaluating the quality of the steganographic image in relation to the cover image. The technique involves calculating the mean of the squared differences between corresponding pixels or patterns in the original and hidden images.

\begin{equation}
    MSE = \sum_{i=0}^{m-1}\sum_{j=0}^{n-1}\left[ \mathbb{X}_c(i, j) - \mathbb{X}_s(i, j)\right]^2
\end{equation}

where $(i, j)$ represents the $i^{th}$ and $j^{th}$ pixel of the cover $\mathcal{X}_c(i, j)$ and steganographic $\mathcal{X}_s(i, j)$ image, and $|m| = |n|$ as the sum of the host's and the steganographic photo's pixels is the same.
\\

The equation representing the PSNR may be stated as follows:
\begin{equation*}
    PSNR = 20\times \log_{10} \left(\dfrac{MAX_C}{\sqrt{MSE}}\right)
\end{equation*}
\begin{equation*}
    PSNR = 20 \times [\log_{10}\left(MAX_C\right) - \log_{10}\left(\sqrt(MSE)\right)]
\end{equation*}  

\begin{equation}
    PSNR = 20 \times \log_{10}\left(MAX_C\right) - 10 \times \log_{10}\left(MSE\right)
\end{equation}  

The variable $MAX_C$ denotes the highest pixel value included in the cover image. Hence, a comprehensive fitness function may be defined as
\begin{equation*}
    Z(\mathbb{X}_s, \mathbb{X}_c)     = \eta \times SSIM(\mathbb{X}_s, \mathbb{X}_c) + (1 - \eta) \times \dfrac{PSNR(\mathbb{X}_s, \mathbb{X}_c)}{100}
\end{equation*}
In this context, $\mathbb{X}_s$ and $\mathbb{X}_c$ denote two images that are being subjected to a comparative analysis. In the context of our study, the symbol $\mathbb{X}_s$ is used to denote a steganographic image, whereas $\mathbb{X}_c$ is used to represent a cover image. Since the cover image $\mathbb{X}_c$ remains unchanged during the embedding process, the fitness function may be expressed as
\begin{equation}\label{eq:objFun}
    Z(\mathbb{S}, \mathbb{C})     = \eta \times SSIM(\mathbb{S}, \mathbb{C}) + (1 - \eta) \times \dfrac{PSNR(\mathbb{S}, \mathbb{C})}{100}
\end{equation}
Here the value of $\eta$ is taken as 0.5.

The Structural Similarity Index (SSIM) value gives a consistent metric for evaluating the quality of a picture. Mathematically, it is defined as:
\begin{equation*}
    Q = \dfrac{4 \sigma_{uv} \hat{u}\hat{v}}{ (\hat{u}^2 + \hat{v}^2) (\sigma_{u}^2 + \sigma_{v}^2) }
\end{equation*}
The window size of the steganographic image is denoted by variables $u$ and $v$, while its average is denoted by variables $\hat{u}$ and $\hat{v}$.
The variance of variables $u$ and $v$ is denoted as $\sigma^u_2$ and $\sigma^v_2$, respectively, whereas their covariance is denoted as $\sigma_{uv}$.

This statistic considers the influence of the Human Visual System (HVS). The model distortions of the HVS consist of three primary elements: luminance distortion, loss of correction, and contrast distortion.
The mathematical models used to represent them are as follows:
\begin{equation}
   Q = \dfrac{2 \sigma_{uv}}{\sigma_{u} \sigma_{v}} \times  \dfrac{2\hat{u}\hat{v}}{\hat{u}^2 + \hat{v}^2} \times  \dfrac{\sigma_{u}\sigma_{v}}{(\sigma_{u}^2 + \sigma_{v}^2)}
\end{equation}

The primary component has a linear relationship and encompasses a comprehensive range of values ranging from -1 to 1. The second element demonstrates a dynamic range including the values of 0 and 1, inclusively. The aforementioned range is used for the purpose of representing the luminance values of variables $u$ and $v$. The third component exhibits a contrast similarity measure that spans from 0 to 1.
Hence, the variable $Q$ is constrained inside the interval [-1, 1]. It is essential to recognize that this condition is true only in cases when the denominator is not equal to zero. When the denominator term in the quotient being examined approaches zero, the expression demonstrates instability. To address this concern, the use of the SSIM is often employed as a means of quantifying the level of similarity between two pictures. The metric is derived by assessing the given equation.
\begin{equation}
    SSIM (u,v) =  \dfrac{(2\hat{u}\hat{v} + c_1)(2\sigma_{uv} + c_2)}{(\hat{u}^2 + \hat{v}^2 + 1) (\sigma_{u}^2 + \sigma_{v}^2 + c_2)}
\end{equation}

Take the window sizes of the original and steganographic images to be $u$ and $v$, respectively. Let the median values of $u$ and $v$ also stand for attention to detail. The symbols $\sigma^u_2$ and $\sigma^v_2$ represent the variances of the variables $u$ and $v$, respectively. On the other hand, $\sigma_uv$ denotes the covariance between $u$ and $v$.

\begin{sidewaysfigure}
    \centering
\includegraphics[width=\linewidth]{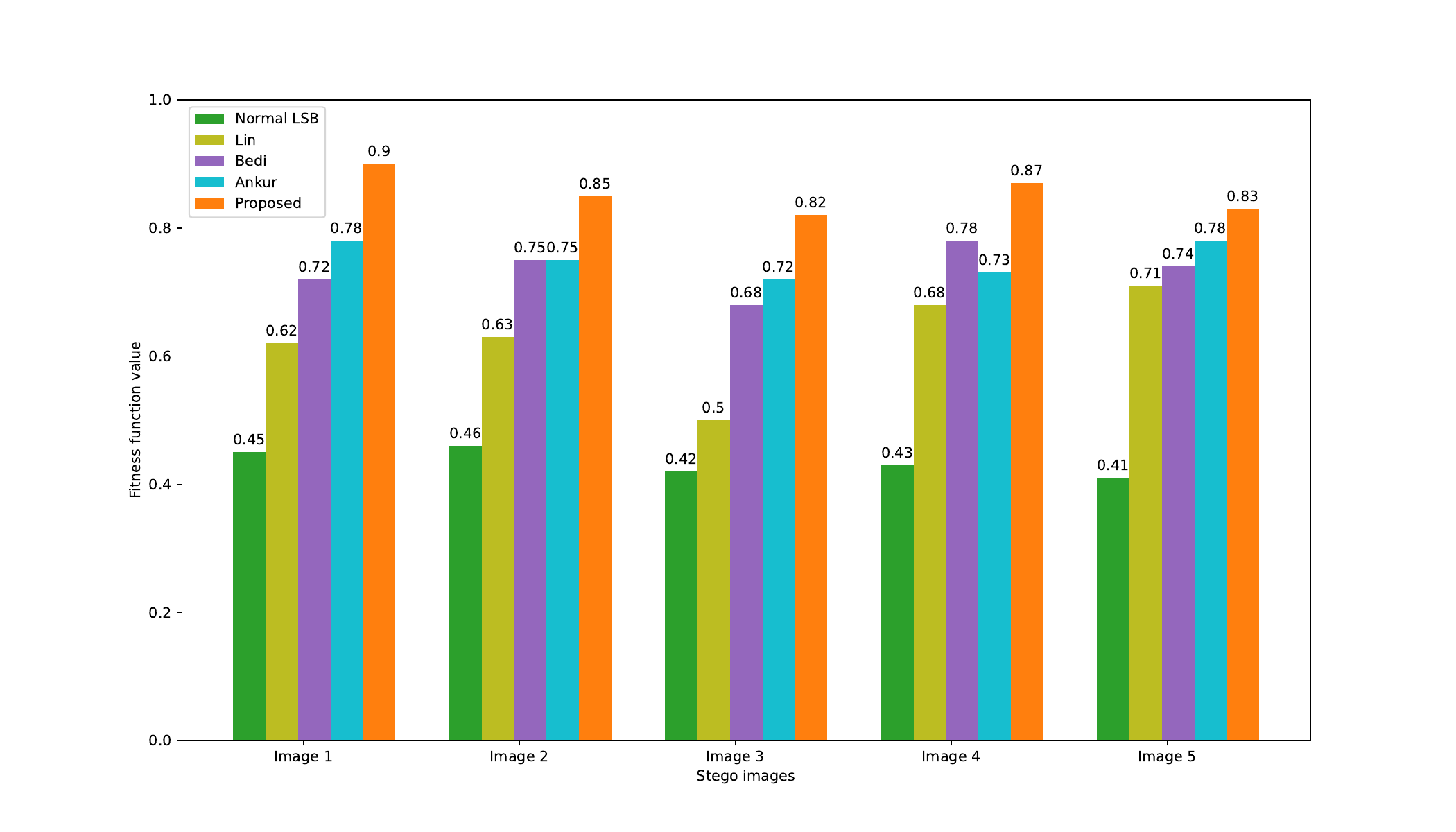}
    \caption{Comparing results of different SOTA steganographic methods}
\end{sidewaysfigure}

\section{Experimental setup \& Results} \label{sec:results}
In the conducted experiments, the starting population is produced using a process of uniformly distributed initialization at random within the specified ranges or limitations of the intended variables. The number of individuals has been established as 40, whereas the size of the benchmark functions is 30. The halting condition for the algorithm was defined as a maximum number of iterations, which was set at 2000. The studies were conducted with a total of 30 unique runs for every procedure and algorithm, using distinct beginning settings for each individual example.

In the classic firefly technique, the starting attractiveness is determined by setting $\beta_{0}$ as $2 * rand$. Additionally, the light absorption coefficient $\gamma$ is calculated as $1/S^2$, where $S$ represents the mean distance of the variables. The random variable $\eta$ is defined as $\eta = \eta_0 \times 0.95^{iter}$, where $\eta_{0} = 0.2$ represents the starting unpredictability component and iter is the iteration index. The reduction occurs in a slow and monotonous manner.

The Lévy distribution is used in order to generate random numbers due to its infrequent production of large jumps, hence decreasing the likelihood of being trapped in a local optima. The differential evolution method employs parameter values of $F = 0.5$ for the coefficient of scaling and $Cr = 0.9$ for the crossover constant. The findings are presented in the next section.

\begin{figure}[ht]
    \centering
    \begin{subfigure}[h]{0.47\linewidth}
        \includegraphics[width=\linewidth]{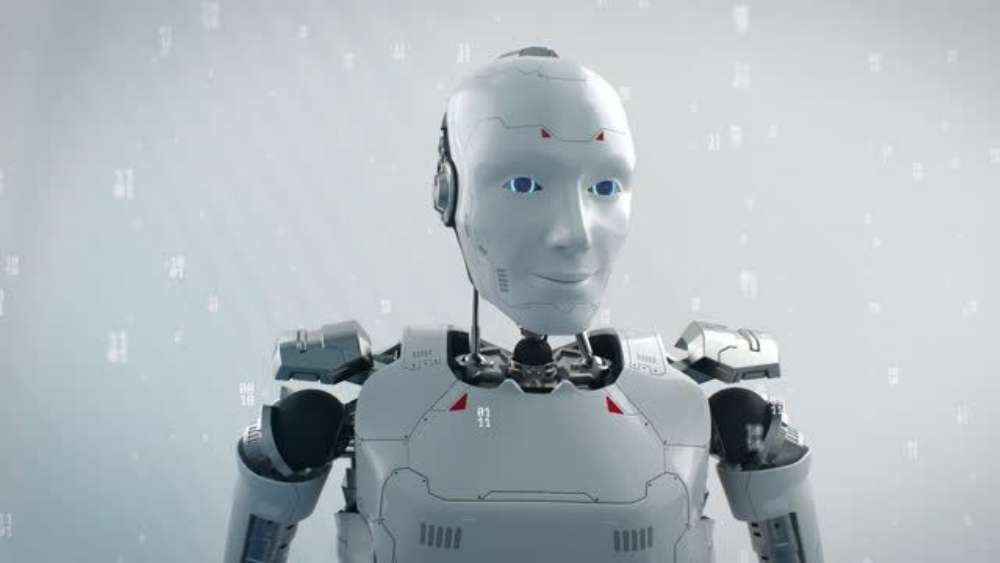}
        \caption{Pre-Steganography}
    \end{subfigure}
    \hfill
    \begin{subfigure}[h]{0.47\linewidth}
        \includegraphics[width=\linewidth]{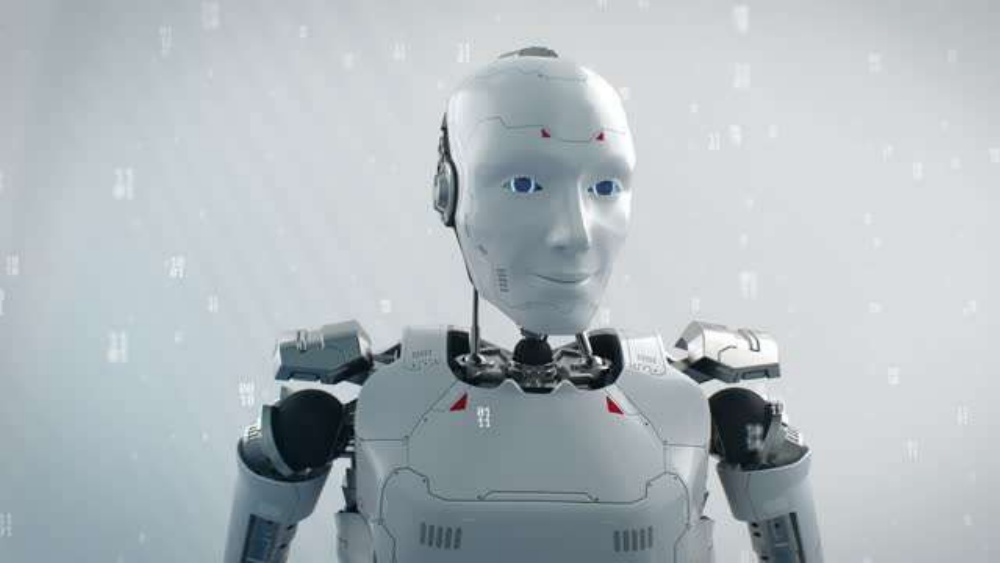}
        \caption{Post-Steganography}
    \end{subfigure}
    \caption{Comparing Pre- and Post-Steganography}
    
\end{figure}

\begin{figure}[ht]%
\begin{subfigure}[h]{0.47\linewidth}
        \includegraphics[width=\linewidth]{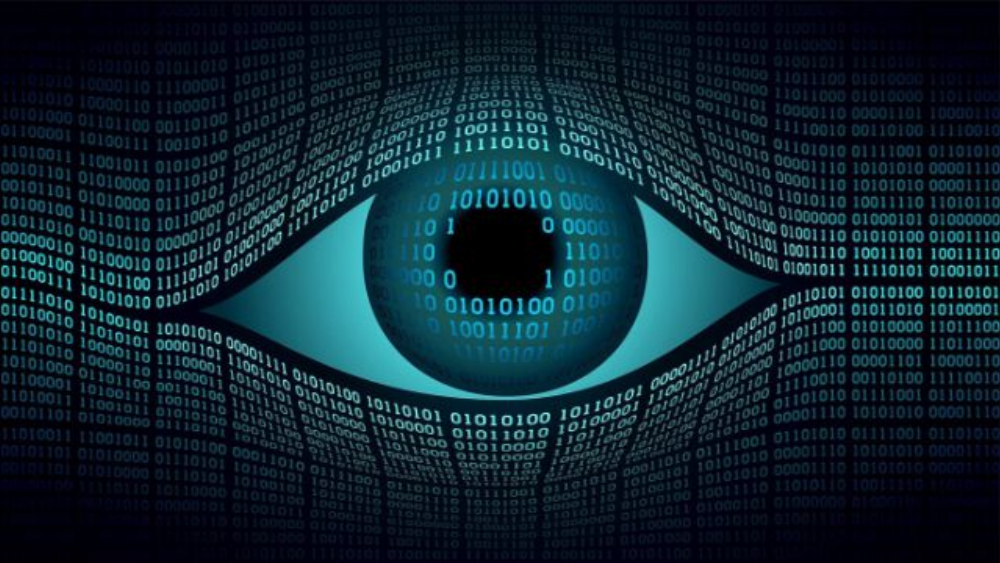}
        \caption{Pre-Steganography}
    \end{subfigure}
    \hfill
    \begin{subfigure}[h]{0.47\linewidth}
        \includegraphics[width=\linewidth]{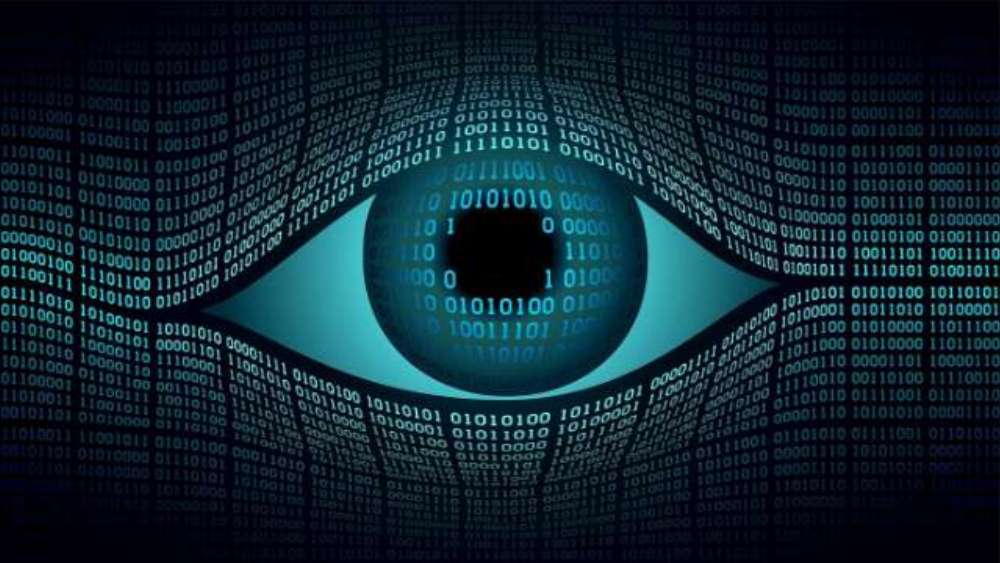}
        \caption{Post-Steganography}
    \end{subfigure}
    \caption{Comparing Pre- and Post-Steganography}
\end{figure}

\begin{figure}[ht]%
    \centering
   \begin{subfigure}[h]{0.47\linewidth}
        \includegraphics[width=\linewidth]{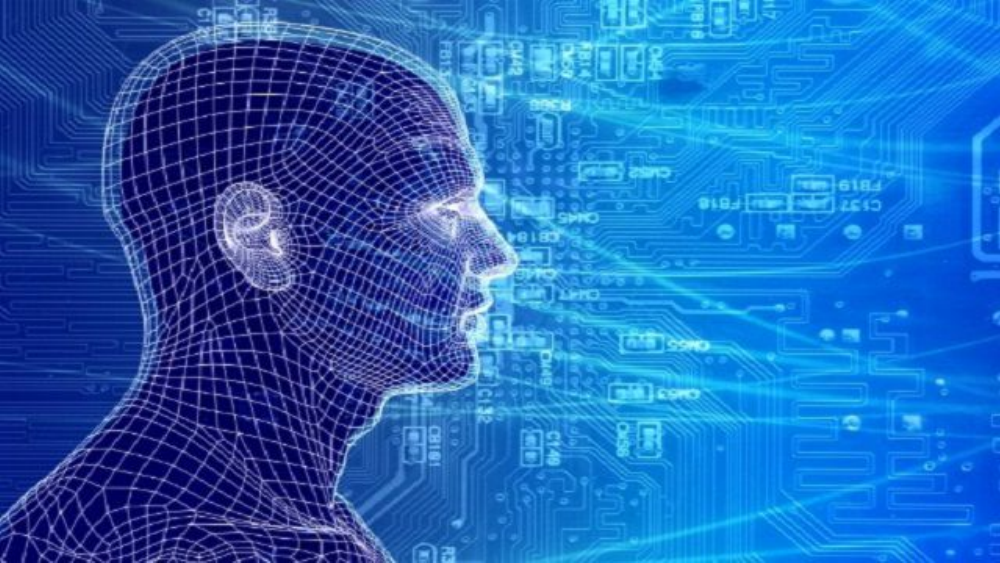}
        \caption{Pre-Steganography}
    \end{subfigure}
    \hfill
    \begin{subfigure}[h]{0.47\linewidth}
        \includegraphics[width=\linewidth]{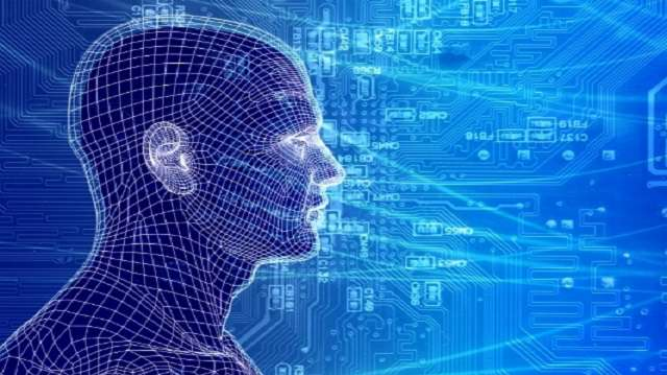}
        \caption{Post-Steganography}
    \end{subfigure}
    \caption{Comparing Pre- and Post-Steganography}
\end{figure}

\begin{figure}[ht]%
    \centering
    \begin{subfigure}[h]{0.47\linewidth}
        \includegraphics[width=\linewidth]{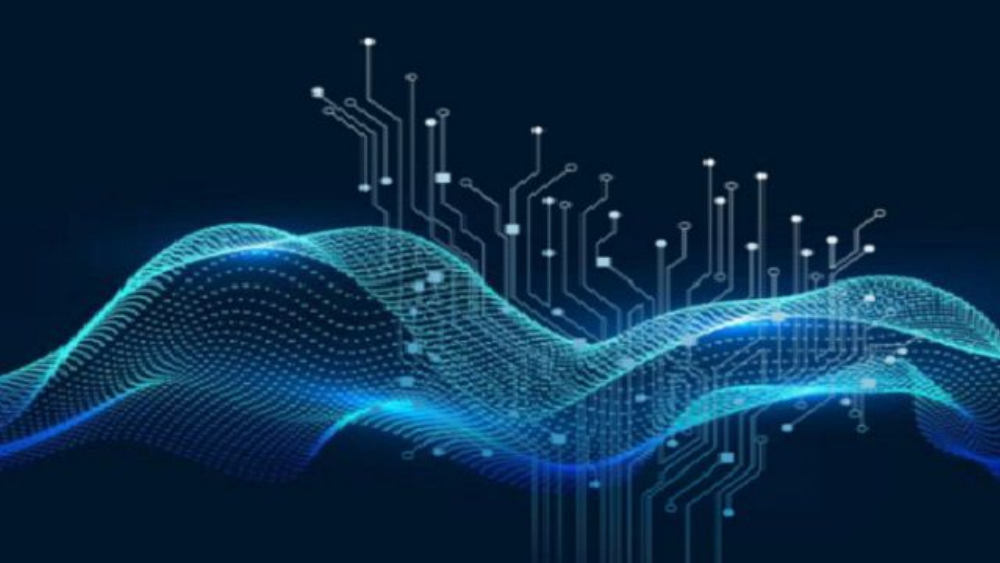}
        \caption{Pre-Steganography}
    \end{subfigure}
    \hfill
    \begin{subfigure}[h]{0.47\linewidth}
        \includegraphics[width=\linewidth]{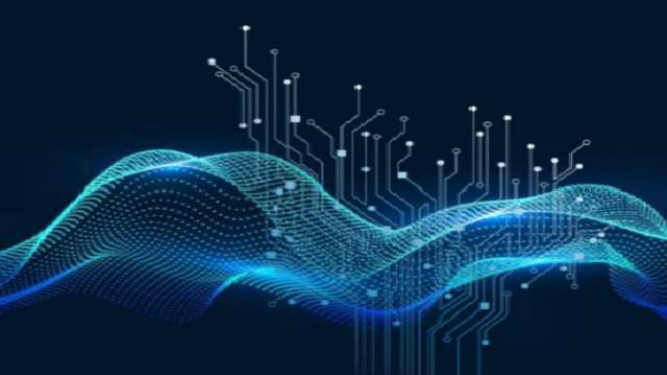}
        \caption{Post-Steganography}
    \end{subfigure}
    \caption{Comparing Pre- and Post-Steganography}
\end{figure}

\begin{figure}[ht]%
    \centering
    \begin{subfigure}[h]{0.47\linewidth}
        \includegraphics[width=\linewidth]{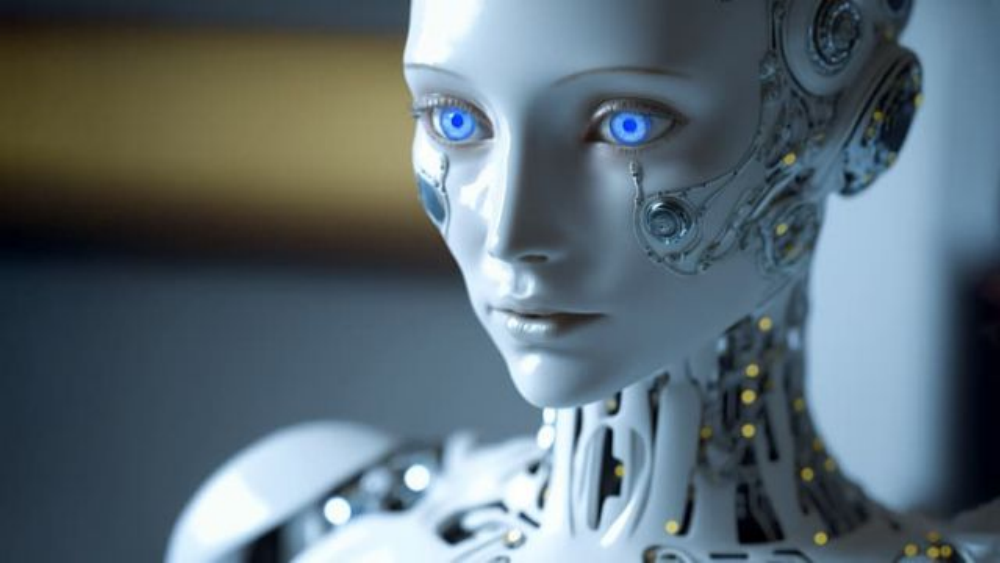}
        \caption{Pre-Steganography}
    \end{subfigure}
    \hfill
    \begin{subfigure}[h]{0.47\linewidth}
        \includegraphics[width=\linewidth]{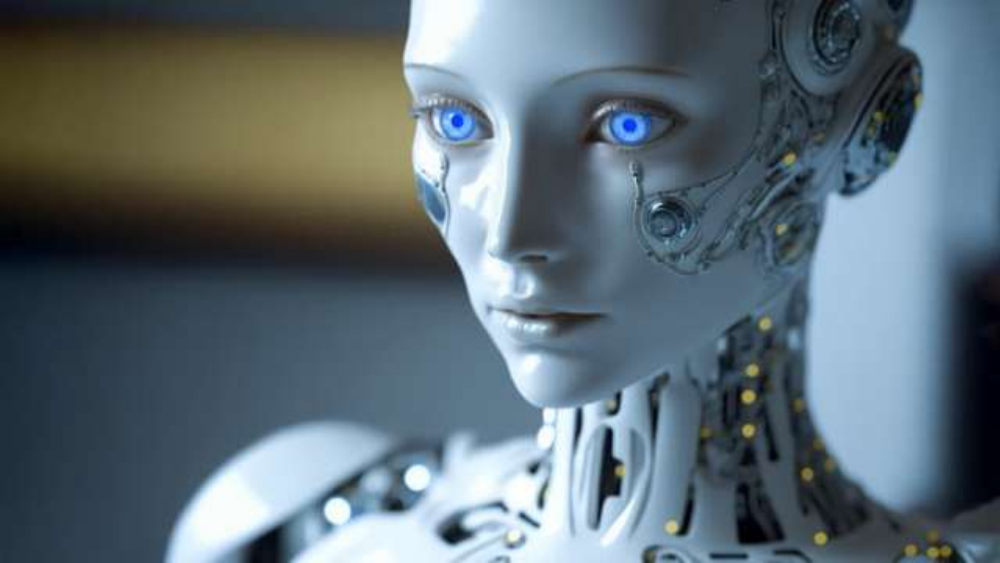}
        \caption{Post-Steganography}
    \end{subfigure}
    \caption{Comparing Pre- and Post-Steganography}
\end{figure}

% \pagebreak
\section{Conclusion} \label{sec:conclusion}
This work presents an innovative methodology for embedding PDF document in an host image, with the objective of minimizing the degradation of image fidelity. The Hybrid Firefly algorithm is employed to get optimal results in terms of minimizing distortion and maximizing efficiency in the process of embedding. The use of this approach successfully addresses the problem of image size inflation that arises from the embedding of PDF file. In the specific context of concealing PDF data inside images without causing noticeable distortion, our methodology exhibits exceptional efficacy in comparison to current cutting-edge methodologies.

\section*{Conflict of Interest} \label{sec:conflict}
The authors of this paper assert that they possess no conflicts of interest.

\section*{Data Availability} \label{sec:dataAvail}
The dataset employed in this work has been acquired from publically accessible sources and is available upon request from the primary author.

% \section*{Ethical approval} \label{sec:ethic}
% This article does not contain any studies with human participants or animals performed by any of the authors.


\bibliography{bibliography}% common bib file

\end{document}